\documentclass[12pt]{article}
\usepackage{graphicx}
\def\beq{\begin{equation}}
\def\eeq{\end{equation}}
\def\bea{\begin{eqnarray}}
\def\eea{\end{eqnarray}}
\def\nn{\nonumber}
\def\sss{\scriptscriptstyle}
\def\lft{{\scriptscriptstyle L}}
\def\rht{{\scriptscriptstyle R}}
\def\roughly#1{\mathrel{\raise.3ex\hbox
{$#1$\kern-.75em\lower1ex\hbox{$\sim$}}}}

\def\sla#1{\raise.15ex\hbox{$/$}\kern-.57em #1}

\def\ks{K_{\sss S}}

\begin{document}

\begin{flushright}
UMiss-HEP-2006-017 \\
UdeM-GPP-TH-06-152 \\
\end{flushright}

\begin{center}
\bigskip
{\Large \bf \boldmath CP Violation in Hadronic $\tau$ Decays} \\
\bigskip
\bigskip
{\large 
Alakabha Datta $^{a,}$\footnote{datta@phy.olemiss.edu},
Ken Kiers $^{b,}$\footnote{knkiers@taylor.edu},
David London $^{c,}$\footnote{london@lps.umontreal.ca}, \\
Patrick J. O'Donnell $^{d,}$\footnote{odonnell@physics.utoronto.ca},
and Alejandro Szynkman $^{c,}$\footnote{szynkman@lps.umontreal.ca}
}
\end{center}

\begin{flushleft}
~~~~~~~~~~~$a$: {\it Dept of Physics and Astronomy, 108 Lewis Hall,}\\
~~~~~~~~~~~~~~~{\it University of Mississippi, Oxford, MS 38677-1848, USA}\\
~~~~~~~~~~~$b$: {\it Physics Department, Taylor University,}\\
~~~~~~~~~~~~~~~{\it 236 West Reade Ave., Upland, Indiana, 46989, USA}\\
~~~~~~~~~~~$c$: {\it Physique des Particules, Universit\'e
de Montr\'eal,}\\
~~~~~~~~~~~~~~~{\it C.P. 6128, succ. centre-ville, Montr\'eal, QC,
Canada H3C 3J7}\\
~~~~~~~~~~~$d$: {\it Department of Physics, University of Toronto,}\\
~~~~~~~~~~~~~~~{\it 60 St.\ George Street, Toronto, ON, Canada M5S
1A7}\\
\end{flushleft}

\begin{center}
\bigskip (\today)
\vskip0.5cm {\Large Abstract\\} \vskip3truemm
\parbox[t]{\textwidth}{We re-examine CP violation in the $\Delta S =
0$ decays $\tau \to N\pi \nu_\tau$ ($N=2,3,4$). We assume that the new
physics (NP) is a charged Higgs. We show that there is no NP
contribution to $\tau \to \pi\pi \nu_\tau$, which means that no CP
violation is expected in this decay. On the other hand, NP can
contribute to $\tau \to N\pi \nu_\tau$ ($N=3,4$). These are dominated
by the intermediate resonant decays $\tau \to \omega \pi \nu_\tau$,
$\tau \to \rho\pi\nu_\tau$ and $\tau \to a_1\pi \nu_\tau$. We show
that the only sizeable CP-violating effects which are possible are in
$\tau \to a_1\pi \nu_\tau \to 4\pi \nu_\tau$ (polarization-dependent
rate asymmetry) and $\tau \to \omega\pi \nu_\tau$ (triple-product
asymmetry).}
\end{center}

\thispagestyle{empty}
\newpage
\setcounter{page}{1}
\baselineskip=14pt

\section{Introduction}

One of the most important problems in particle physics is finding the
origin of CP violation. CP violation was first seen in the kaon system
about 40 years ago and was observed in $B$ decays some 5 years ago
\cite{pdg}. If one hopes to understand CP violation, it is important
to look for its effects in as many decays as possible. One of the
purposes of the present paper is to push for the search for CP
violation in hadronic $\tau$ decays. This is not a new area of
research -- there is already a large literature on the subject
\cite{CLS,Kpifinaltheory,3pifirst,3pifinal,decker87,deckermirkes,wise,KM1992,KM1997,Kuhn:1993ra}.
However, there have been few experimental results lately, and many
studies have focused on decays which are less likely to exhibit CP
violation such as $\tau\to\pi\pi\nu_\tau$ \cite{CPVpipiexpt}.

All CP-violating effects come about due to the interference of (at
least) two amplitudes with a relative weak (CP-odd) phase. In the
standard model (SM) all $\tau$ decays involve (essentially) only a
virtual $W$. To the extent that there is only one decay amplitude, all
CP violation in $\tau$ decays vanishes in the SM. That is, the
observation of CP violation in $\tau$ decays would be a smoking-gun
signal of physics beyond the SM. (Note that there can be a SM CP
asymmetry in $\tau \to \pi \ks \nu_\tau$ due to the several amplitudes
in the $\ks$.  The CP asymmetry is $O(10^{-3})$, see Ref.~\cite{BS}.)

Such CP violation will necessarily involve the interference of the SM
$W$ diagram with some new-physics (NP) amplitude. This NP amplitude
typically consists of a right-handed $W_{\sss R}$ or a charged-Higgs
exchange. We show that $W$-$W_{\sss R}$ interference always leads to
effects which are negligible (they are suppressed by $m_\nu$). Thus we
need only consider a NP charged-Higgs exchange. This assumption, while
not completely model-independent (e.g.\ we do not consider light
leptoquarks), is fairly general -- a great many models of NP include a
light charged Higgs boson.

This said, many such models have two Higgs doublets, which give mass
to the fermions. In such models, the coupling of the charged Higgs
boson to first-generation final states is generally proportional to
$m_u$ and $m_d$, which are tiny. As a result, any CP violation which
is due to charged-Higgs exchange is correspondingly small. If CP
violation is to be observed in $\tau$ decays, the charged-Higgs
coupling must be large. Thus, $\tau$ decays probe non-``standard'' NP
CP violation. This is arguably less likely to be observed, but if
seen, the payoff is huge.

In this paper we focus on the decays $\tau \to N\pi \nu_\tau$
($N=2,3,4$). We first show that the charged Higgs cannot contribute to
$\tau \to \pi\pi \nu_\tau$, so that no CP violation is expected in
these decays. By contrast, the charged Higgs can contribute to decays
with three and four pions, so that CP violation is possible. These
decays are dominated by intermediate resonances (mostly vector). A
second purpose of the present paper is to argue that, in studying
$\tau \to N\pi \nu_\tau$ ($N=3,4$), it is best to concentrate on CP
violation in the resonant decays. To this end, we present a study of
CP-violating effects in $\tau \to V\pi \nu_\tau$
($V=\omega,\rho,a_1$). We find that the total rate asymmetries in all
decays are tiny. The polarization-dependent rate asymmetry may be
measurable in $\tau \to a_1\pi \nu_\tau$. A triple-product asymmetry
may be measured in $\tau \to \omega\pi \nu_\tau$. This involves the
measurement of the spin of the $\omega$, which in turn implies
knowledge of the $\tau$ momentum. We do not expect to measure large
CP-violating effects in $\tau \to \rho\pi \nu_\tau \to 3\pi \nu_\tau$.

Sec.~2 contains a proof that $W$-$W_{\sss R}$ CP-violating effects are
negligible, which leads us to consider charged-Higgs exchange as the
new physics. We then discuss three possible CP-violating effects that
are allowed, in general: the total rate asymmetry, the
polarization-dependent rate asymmetry, and the triple-product
asymmetry. We then examine the charged-Higgs contribution to $\tau \to
N\pi \nu_\tau$ ($N=2,3,4$). Our main results are found in Sec.~3. We
consider resonant contributions to the hadronic $\tau$ decays and
examine $\tau \to \omega \pi \nu_\tau$, $\tau \to \rho\pi\nu_\tau$ and
$\tau \to a_1\pi \nu_\tau$. We write the form factors for the decays
$\tau \to V\pi \nu_\tau$ ($V=\omega,\rho,a_1$) and examine the
possible CP-violating effects. We find that such effects are possible
in $\tau \to a_1\pi \nu_\tau \to 4\pi \nu_\tau$
(polarization-dependent rate asymmetry) and $\tau \to \omega\pi
\nu_\tau$ (triple-product asymmetry). We conclude in Sec.~4. Several
technical details are included in the Appendix.

\section{$\tau$ Decays: Generalities}

\subsection{SM-NP Interference}

Consider the decay $\tau \to f \nu_\tau$, where $f$ is some final
hadronic state. Assume also that the SM and NP contribute to this
decay. Both contributions can be split into a leptonic and a hadronic
piece. The SM leptonic piece is proportional to ${\bar u}_\nu \gamma_\mu
\gamma_\lft u_\tau$, where $\gamma_\lft = (1- \gamma_5)/2$.

Now suppose that the NP consists of a right-handed $W_{\sss R}$. In
this case, the NP leptonic piece is proportional to ${\bar u}_\nu
\gamma_\mu \gamma_\rht u_\tau$, where $\gamma_\rht = (1+ \gamma_5)/2$.
The SM-NP interference thus includes a piece proportional to ${\rm Tr}
[ u_\nu {\bar u}_\nu \gamma_\mu \gamma_\lft u_\tau {\bar u}_\tau
\gamma_\nu \gamma_\rht ]$. This trace is proportional to $m_\nu$:
$\gamma_\rht u_\nu {\bar u}_\nu \gamma_\rht \sim m_\nu$. Since $m_\nu$
is negligible, we therefore conclude that CP violation in $\tau$
decays coming from $W$-$W_{\sss R}$ interference is negligible, and
take the NP to consist only of a charged Higgs boson $H$.

The coupling of the charged Higgs to leptons is scalar and/or
pseudoscalar. This coupling can be written in terms of $\gamma_\lft$
and $\gamma_\rht$. The piece of the neutrino coupling proportional to
$\gamma_\rht$ leads to a $W$-$H$ interference term which is
proportional to $m_\nu$ (as in the interference of $W$ and $W_{\sss
R}$ above), and is negligible. Thus, it is only the $H$-neutrino
coupling proportional to $\gamma_\lft$ which is important. The $W$-$H$
interference term then involves
\bea
{\rm Tr} [ u_\nu {\bar u}_\nu \gamma_\mu \gamma_\lft u_\tau {\bar
u}_\tau \gamma_\lft ] & = & {\rm Tr} \left[ \sla{p}_\nu \gamma_\mu
\gamma_\lft (\sla{p}_\tau + m_\tau ) {(1 + \gamma_5 \sla{s}_\tau )
\over 2} \gamma_\lft \right] \nn\\
& = & \frac12 m_\tau {\rm Tr} [ \sla{p}_\nu \gamma_\mu \gamma_\lft ] +
\frac12 {\rm Tr} [ \sla{p}_\nu \gamma_\mu \sla{p}_\tau \sla{s}_\tau
\gamma_\lft ] ~.
\label{leptonic}
\eea
The first term is always present; the second term is only important if
the spin of the $\tau$ is known (i.e.\ if the $\tau$ is polarized).

\subsection{CP-Violating Effects}

If there are two contributing amplitudes to the decay $\tau \to f
\nu_\tau$ then the full decay amplitude can be written
\beq{\cal A} = A_1 + A_2 e^{i\phi} e^{i\delta} ~,
\eeq
where $\phi$ and $\delta$ are the relative weak (CP-odd) and strong
(CP-even) phases, respectively. The full rate is proportional to
$\sum_{spins} |{\cal A}|^2$. If some spins are measured, one does not
sum over them.

One CP-violating signal occurs if the spin-independent or spin-dependent
rate is different from that of its anti-process. This is known as
direct CP violation. It is proportional to 
\beq 
\sin \phi \sin \delta ~.
\eeq
Thus, one can obtain a direct CP asymmetry if the two decay amplitudes
have a nonzero relative weak {\it and} strong phase.

A second signal involves a {\it triple product} (TP). The TP takes the
form $\vec v_1 \cdot (\vec v_2 \times \vec v_3)$, where each $v_i$ is
a spin or momentum. In a 2-body $\tau$ decay, the TP must involve 2
spins -- since the $\nu_\tau$ cannot be measured, this must include
the spin of the $\tau$. The TP for a 3-body decay includes 1 spin; all
the $v_i$ can be momenta in an $M$-body ($M\ge 4$) decay. If the TP in
a given $\tau$ decay is different from that seen in its anti-process,
this is another signal of CP violation. The TP asymmetry is
proportional to
\beq 
\sin \phi \cos \delta ~.
\eeq
Thus, one does not require a strong-phase difference to get a TP
asymmetry.

As discussed above, all CP-violating effects require the interference
of two amplitudes. In particular, the transition $H \to f$ must be
possible. For example, suppose that $f$ can be produced only through
the decay of a vector (or axial-vector) meson $V$. Since the spin-0
charged Higgs cannot couple to the spin-1 vector meson, the decay
$\tau \to V \nu_\tau \to f \nu_\tau$ can come only from a SM virtual
$W$ exchange.  In this case there is only a single amplitude and one
will not have CP violation. The lesson here is that one must check
whether the final-state $f$ can be produced from a spin-0 charged
Higgs boson.

\subsection{\boldmath $\Delta S = 0$ $\tau$ Decays: $\tau \to N\pi
\nu_\tau$}

Above, we argued that, in the decay $\tau \to f \nu_\tau$, it is
crucial to know if the transition $H\to f$ is allowed. Here we
consider $\tau \to N\pi \nu_\tau$, with $N=2,3,4$. For a given $N\pi$
final hadronic state one can use isospin to determine if $H\to f$ can
take place.

Consider $N=2$, i.e.\ $\tau \to \pi \pi \nu_\tau$. The intermediate
particle ($W$, $H$) decays to ${\bar u} d$. The $\pi\pi$ final state
can be produced if a gluon decays to a $u{\bar u}$ or $d{\bar d}$
pair, and these quarks combine with the ${\bar u} d$ from the decaying
particle. In the isospin limit, the $u{\bar u}$ and $d{\bar d}$ quark
pairs have the same coupling to the gluons, and are produced with the
same rate. The amplitude for $\tau^- \to \pi^- \pi^0 \nu_\tau$ thus
involves the formation of mesons from quarks.  Under isospin we have
$\pi^+ \equiv \bar d u$, $\pi^- \equiv - \bar u d$ and $\pi^0 \equiv
(\bar d d - \bar u u)/\sqrt{2}$.  Thus,
\beq
{\bar u} (u {\bar u}) d + {\bar u} (d {\bar d}) d \sim [\pi^0 \pi^- -
\pi^- \pi^0]~. 
\label{antisospin}
\eeq
\textit{A priori}, the state of two pions corresponds to total isospin
$I=0,1,2$. However, this antisymmetric combination corresponds to
$I=1$ only. (Note: the total isospin must be $I=1$ because the gluons,
which produce the quark pairs, have $I=0$.) Since the total wave
function of two pions must be symmetric (Bose symmetry), the relative
orbital angular momentum must be odd, giving a spatial wave function
which is also antisymmetric. This says that the $\pi\pi$ final state
{\it cannot} be produced from a spin-0 charged Higgs boson. Thus, no
CP violation is expected in this decay (more precisely, it is only at
the level of isospin breaking). This is what is found experimentally
\cite{CPVpipiexpt}.

Note that the above argument does not hold when the final state $A^-
B^0$ does not involve identical particles. Examples of this are
$\rho^0 \pi^- $, $a_1^- \pi^0$, $\omega^0 \pi^- $, etc. That is, we
know that the final state must have $I=1$, but this does not imply
anything about the total angular momentum in the absence of identical
particles. Thus, $J=0$ is allowed, and the final state $A^- B^0$ can
be produced in the decay of a charged Higgs. For these final states,
CP-violating effects are expected.

A special case of this is the decay $\tau\to \pi K\nu_\tau$ which has
been considered in Refs.~\cite{CLS,Kpifinaltheory,BS,CPVKpiexpt}. Here
isospin gives $K^- \equiv - \bar u s$ and ${\bar K}^0 \equiv \bar d
s$. Thus,
\beq
{\bar u} (u {\bar u}) s + {\bar u} (d {\bar d}) s \sim \left[ (1/2)
\pi^0 K^- - \pi^- {\bar K}^0 \right]~.
\eeq
Here the two pieces do not contain the same particles, and so the
argument used above for $\tau \to \pi \pi \nu_\tau$ does not apply.
Thus, the decay $\tau\to \pi K\nu_\tau$ receives (non-resonant)
contributions from a charged Higgs boson. In addition, $\tau \to \pi K
\nu_\tau$ can also be produced resonantly, from an intermediate vector
$V$ [$K^*$(892)] or an intermediate scalar $S$ [$K^*_0$(1430)]
\cite{CLS}. Note that the charged Higgs can couple to an $S$. Thus CP
violation is possible in $\tau\to \pi K\nu_\tau$. Since we have
assumed that the coupling of the charged Higgs is arbitrary -- that
is, it is not proportional to quark masses -- this CP violation could
be large. However, it is also possible that the NP follows the pattern
of the CKM matrix, so that CP violation in $\Delta S = 1$ decays is
suppressed compared to that in $\Delta S = 0$ decays. In this paper we
focus on $\Delta S = 0$ decays of the $\tau$.

Now consider $N=3$: $\tau \to \pi \pi \pi \nu_\tau$
\cite{3pifirst,3pifinal}. This actually represents two decays: $\tau^-
\to \pi^- \pi^+ \pi^- \nu_\tau$ (called $0\pi^0$) and $\tau^- \to
\pi^- \pi^0 \pi^0 \nu_\tau$ ($2\pi^0$). We write,
\bea
\pi\pi\pi & \sim & {\bar u} (u {\bar u}) (u {\bar u}) d + {\bar u} (d {\bar
d}) (d {\bar d}) d + {\bar u} (u {\bar u}) (d {\bar d}) d + {\bar u}
(d {\bar d}) (u {\bar u}) d \nn\\
& \sim & -\pi^0 \pi^0 \pi^- -\pi^- \pi^0 \pi^0 +\pi^0 \pi^- \pi^0
+\pi^- \pi^+ \pi^- ~.
\eea
In this case there is no constraint on the total angular momentum of
the $0\pi^0$ or $2\pi^0$ state, and the spin-0 $H$ can contribute
directly to both $3\pi$ $\tau$ decays in addition to the spin-1 $W$.

The $N=4$ decay is similar: $\tau \to \pi \pi \pi \pi \nu_\tau$. As
above, this actually represents two decays: $\tau^- \to \pi^- \pi^+
\pi^- \pi^0 \nu_\tau$ ($1\pi^0$) and $\tau^- \to \pi^- \pi^0 \pi^0
\pi^0 \nu_\tau$ ($3\pi^0$). Given that the additional quarks come from
gluons,
\bea
\pi\pi\pi\pi & \sim & {\bar u} (u {\bar u}) (u {\bar u}) (u {\bar u})
d + {\bar u} (u {\bar u}) (u {\bar u}) (d {\bar d}) d + {\bar u} (u
{\bar u}) (d {\bar d}) (d {\bar d}) d \nn\\
& & \hskip0.4truein +~{\bar u} (d {\bar d}) (d {\bar d}) (d {\bar d})
d + {\bar u} (d {\bar d}) (u {\bar u}) (d {\bar d}) d + {\bar u} (d
{\bar d}) (u {\bar u}) (u {\bar u}) d \nn\\
& & \hskip0.4truein +~{\bar u} (d {\bar d}) (d {\bar d}) (u {\bar u})
d + {\bar u} (u {\bar u}) (d {\bar d}) (u {\bar u}) d \nn\\
& \sim & \pi^0 \pi^0 \pi^0 \pi^- - \pi^0 \pi^0 \pi^- \pi^0 + \pi^0
\pi^- \pi^0 \pi^0 - \pi^- \pi^0 \pi^0 \pi^0 \nn\\
& & \hskip0.4truein +~\pi^- \pi^+ \pi^- \pi^0 - \pi^- \pi^+ \pi^0
\pi^- + \pi^- \pi^0 \pi^+ \pi^- - \pi^0 \pi^- \pi^+ \pi^- ~.
\eea
The first and last four terms contribute to the decays $3\pi^0$ and
$1\pi^0$, respectively. These terms do not have any particular symmetry
under the exchange of any two pairs of $\pi$'s, so that there is no
constraint on the total angular momentum of the final state. Thus, the
spin-0 $H$ can contribute directly (non-resonantly) to both $4\pi$
$\tau$ decays.

\section{\boldmath $\tau \to \omega \pi 
\nu_\tau$/$\rho\pi\nu_\tau$/$a_1\pi \nu_\tau$}

We have therefore seen that there can be non-resonant contributions to
$\tau \to N\pi \nu_\tau$ ($N=3,4$) coming from the charged Higgs.
However, there can also be resonant contributions from intermediate
states such as $\omega\pi$, $\rho\pi$ and $a_1\pi$. These are 3-body
decays, while the non-resonant decays are 4-body or 5-body. The
phase-space suppression of these latter decays implies that the widths
of such decays are much smaller than those of resonant decays, and so
$\tau \to N\pi \nu_\tau$ is dominated by resonances. (This is
consistent with the results of Ref.~\cite{4pi}, which finds that the
$4\pi$'s in $\tau \to \pi \pi \pi \pi \nu_\tau$ come primarily from the
(resonant) decay of $\omega \pi$ or $a_1 \pi$.) In the following we
neglect all non-resonant contributions.

Thus, there are several $\Delta S = 0$ $\tau$ decays which can receive
contributions from a spin-0 charged Higgs.  That is, charged $H$ (NP)
can give rise to CP violation. Although the decay $\tau \to \pi\pi
\nu_\tau$ will not manifest CP violation, such an effect can be seen
in $\tau \to \omega \pi \nu_\tau$, $\tau \to \rho \pi \nu_\tau$, and
$\tau \to a_1 \pi \nu_\tau$. Obviously, the amount of CP violation
will depend on the size of the $H\to f$ contribution.

In this section we examine CP violation in resonant $\Delta S = 0$
decays. There are two ingredients: the size of form factors (for both
the SM and the NP), and the size of the hadronic charged Higgs
coupling. These will be discussed in turn.

The matrix element for $\tau \to f \nu_\tau$ involves the product of a
leptonic and hadronic current for the the $W$ and the charged Higgs.
However, as we saw in Sec.~2.1, the Higgs' leptonic current must
contain a left-handed coupling to neutrinos in order to interfere with
that of the SM.  We can therefore write the Hamiltonian as
\beq
H = \frac{G_F}{\sqrt{2}}\cos\theta_c \left( L_{\mu} H^{\mu} + L_0 H_0
\right) + h.c. ~,
\label{hamiltonian}
\eeq
where the first and second pieces correspond to $W$ and $H$ exchange,
respectively. We have
\bea
L_{\mu} = \bar{\nu}\gamma_\mu(1-\gamma_5)\tau & ~~,~~~~ & H^{\mu} =
\bar{d} \gamma^\mu(1-\gamma_5) u \nn\\
L_{0} = \bar{\nu}(1 + \gamma_5)\tau & ~~,~~~~ & H_{0} = \bar{d}\,\,(a
+ b\gamma_5\,)\,u ~.
\label{current}
\eea
To evaluate the matrix element of $\tau \to f \nu_\tau$ we therefore
have to estimate the hadronic matrix elements $\langle f|H^{\mu}|0
\rangle$ and $\langle f|H_{0}|0 \rangle$.

\subsection{\boldmath $\tau \to \omega \pi 
\nu_\tau$/$\rho\pi\nu_\tau$/$a_1\pi \nu_\tau$: Form Factors}

Consider the decay $\tau(l) \to V(q_1) \pi(q_2) \nu_\tau(l^\prime)$,
where $V$ is a vector or axial-vector meson. The general structure for
the SM current $J^{\mu}=\langle V(q_1)\pi(q_2)|H^\mu|0\rangle$
is \cite{decker87,deckermirkes}
\bea
    J^{\mu} & = & F_1(Q^2)\left(Q^2\epsilon_1^\mu-\epsilon_1\cdot q_2 Q^\mu\right)
       +F_2(Q^2)\,\epsilon_1\cdot q_2
         \left(q_1^\mu-q_2^\mu-Q^\mu\frac{Q\cdot(q_1-q_2)}{Q^2}\right) \nonumber\\
     & & +iF_3(Q^2)\,\varepsilon^{\mu\alpha\beta\gamma}
             \epsilon_{1\alpha}q_{1\beta}q_{2\gamma}
       +F_4(Q^2)\,\epsilon_1\cdot q_2 Q^\mu \; ,
\label{Wamp}
\eea 
where $Q^\mu\equiv (q_1+q_2)^\mu$ and $\epsilon_1$ denotes the
polarization tensor of the $V$.  The latter satisfies the relations
\bea
    q_1\cdot \epsilon_1 &=& 0 \; , \\
    \sum_\lambda\epsilon_1^\mu\epsilon_1^{*\nu} & = & 
       -g^{\mu\nu}+\frac{q_1^\mu q_1^\nu}{m_V^2} \; .
\eea

The SM hadronic current is described by the four form factors
$F_{1{\hbox{-}}4}(Q^2)$. For a vector meson ($\omega$ or $\rho$),
experimentally the contribution proportional to $F_3$ dominates.  At
present there is no evidence for the ``second-class currents''
proportional to $F_1$ and $F_2$~\cite{4pi,aleph,aleph1997} (which
violate $G$-parity). For an axial-vector meson ($a_1$) $F_3$ switches
roles with $F_1$ and $F_2$. That is, for the $a_1$, the contributions
proportional to $F_1$ and $F_2$ dominate, while the $F_3$ term
violates $G$-parity. In both cases the term proportional to $F_4$ can
arise only from a scalar interaction. In the SM, this is negligible
\cite{deckermirkes}.  A significant $F_4$ term can only be produced
from NP.

The current $J^\mu$ can receive pole contributions. For example, for
$\omega\pi$ these come from the $\rho$ and $\rho'=\rho(1450)$, which
contribute to $F_3$,
\bea 
   F_3^{\textrm{\scriptsize{pole}}}(Q^2) & = & 
    \frac{g_{\rho\omega\pi}}{\gamma_\rho}\left[
      \frac{m_\rho^2}{Q^2-m_\rho^2+i\sqrt{Q^2}\Gamma_\rho}+A_1
      \frac{m_{\rho^\prime}^2}
	{Q^2-m_{\rho^\prime}^2+i\sqrt{Q^2}\Gamma_{\rho^\prime}}\right]\; .
\label{omega_res}
\eea
In this expression, $g_{\rho \omega \pi}$ is the $\rho\omega\pi$ 
strong coupling and $\gamma_\rho$ is a quantity associated with the
weak coupling of the $\rho$ to the weak charged current~\cite{4pi}.
The constant $A_1$ may be expressed as a ratio comparing $g_{\rho \omega \pi}$
and $\gamma_\rho$ to the analogous quantities for the $\rho^\prime$.
In Ref.~\cite{4pi}, $A_1$ is taken as a parameter that is fit to the
data.  It is
now clear that the widths of the $\rho$ and the $\rho'$ can generate a
strong phase for the form factors. It should be pointed out that our
model of the resonant contribution is simple and should be understood
as a demonstration of how strong phases can arise. Note that, if
$G$-parity is violated, there can also be resonant contributions to
$F_1(Q^2)$ and $F_2(Q^2)$ \cite{deckermirkes}.

The non-resonant part of $J^{\mu}$ may be estimated in Chiral
Perturbation Theory~\cite{wise} as an expansion in the pion momentum
with the vector mesons treated as heavy. However, in $\tau$ decays the
pion is not always soft and so higher-order terms in the chiral
Lagrangian are important, leading to a loss of predictive power. Hence
we do not estimate the non-resonant part.

We now turn to the charged Higgs' hadronic current ($H_0$). Since the
charged Higgs has spin 0, the $V \pi$ system is in a relative $L=1$
angular momentum state and so only the pseudoscalar (scalar) term in
Eq.~(\ref{current}) can contribute to the hadronic current for a
vector (axial-vector) meson $V$. The general structure is
\bea
J_{\textrm{\scriptsize{Higgs}}} & = & \langle
V(q_1)\pi(q_2)|H_0|0\rangle = 
\cases{
bf_{\sss H} \, \epsilon_1 \cdot q_2, & $V$: vector, \cr
af_{\sss H} \, \epsilon_1 \cdot q_2, & $V$: axial-vector,}
\label{Hamp}
\eea
where $a$ and $b$ are, respectively, the scalar and pseudoscalar
couplings of the charged Higgs [Eq.~(\ref{current})].  The form factor
$f_{\sss H}$ can have a CP-conserving strong phase from rescattering
while $a$ and $b$ will in general have a CP-violating weak phase from
New Physics. 

The squared matrix element is now given as
\bea
    \left|{\cal A}\right|^2 & = & \frac{G_F^2}{2}\cos^2\theta_c 
     \left( M_{\mu} J^{\mu} + M_{\textrm{\scriptsize{Higgs}}} J_{\textrm{\scriptsize{Higgs}}}
     \right)\left( M_{\nu} J^{\nu} 
       + M_{\textrm{\scriptsize{Higgs}}} J_{\textrm{\scriptsize{Higgs}}}
     \right)^{\dagger} \; ,
   \label{eq:Asq}
\eea
where the $M$'s are the matrix elements of the leptonic currents in
Eq.~(\ref{current}). Using the equations of motion for the $\tau$ and
the neutrino, we can rewrite Eq.~(\ref{eq:Asq}) as
\bea
    \left|{\cal A}\right|^2 & = & \frac{G_F^2}{2}\cos^2\theta_c 
     L_{\mu\nu}\widetilde{H}^{\mu\nu} \; ,
   \label{eq:Asq_2}
\eea
where $L_{\mu\nu}\equiv M_\mu\left(M_\nu\right)^\dagger$ and
$\widetilde{H}^{\mu\nu}\equiv
\widetilde{J}^\mu\left(\widetilde{J}^\nu\right)^\dagger$, with
\beq
\widetilde{J}^\mu = 
\cases{
J^\mu + b (f_H/m_\tau) \epsilon_1\cdot q_2 Q^\mu, & $V$: vector, \cr
J^\mu + a (f_H/m_\tau) \epsilon_1\cdot q_2 Q^\mu, & $V$: axial-vector.}
    \label{eq:Jmu2}
\eeq
Comparing Eqs.~(\ref{Wamp}) and (\ref{eq:Jmu2}), we see that
$\widetilde{J}^\mu$ is obtained from $J^\mu$ by the replacement
$F_4(Q^2)\to \widetilde{F}_4(Q^2)$, where
\bea
\widetilde{F}_4(Q^2)=
\cases{
F_4(Q^2)+ ({bf_H}/{m_\tau}), & $V$: vector, \cr
F_4(Q^2)+ ({af_H}/{m_\tau}), & $V$: axial-vector.}
\eea
Since any CP-violating effect requires an interference with the NP, we
immediately see that $\widetilde{F}_4(Q^2)$ must be involved in any
such effect.

For future reference we note that, in the hadronic rest frame (where
$\vec{Q}=0$ and $Q^0=\sqrt{Q^2}$), we have
\bea
    \widetilde{J}^0 & = & \widetilde{F}_4(Q^2)\epsilon_1\cdot q_2 Q^0 \; ,
      \label{eq:J0}\\
    \widetilde{J}^i & = & J^i = F_1(Q^2)\,Q^2\epsilon_1^i+2F_2(Q^2)\,\epsilon_1\cdot q_2 \,q_1^i
          -iF_3(Q^2)\,\varepsilon^{ij0k}\epsilon_{1j}q_{1k}Q_0 \; .
      \label{eq:Ji}
\eea
Thus, in the hadronic rest frame, only $\widetilde{J}^0$ contains a NP contribution.

Now for the decay $\tau^-(l) \to V(q_1) \pi^-(q_2) \nu_{\tau}(l^\prime)$ we
can write the differential decay width as \cite{pdg}
\bea
    d \Gamma & = & \frac{(2 \pi)^4}{2m_{\tau}}|{\cal A}|^2 d \Phi_3 \nonumber\\
    d \Phi_3 &= & \delta^4( l- q_1-q_2-l^\prime) 
    \frac{d^3 q_1}{(2 \pi)^3 2E_{q_1}}
\frac{d^3 q_2}{(2 \pi)^3 2E_{q_2}}
    \frac{d^3 l'}{(2 \pi)^3 2E_{l'}}.\
\eea
A similar differential distribution can be written for the CP
conjugate process $\tau^+(l) \to V(q_1) \pi^+(q_2)
\nu_{\tau}(l^\prime)$. 

The differential distribution is best analyzed in the rest frame of
the hadron ($\vec{q_1} + \vec{q_2} =0 $), as is done in
Refs.~\cite{deckermirkes,KM1992,KM1997}. (In the following we follow
closely the conventions and notation of these references.) We define
the angles $\alpha$, $\beta$, $\theta$ and $\psi$ in the same manner
as they are defined in Ref.~\cite{deckermirkes}.  The definitions are
reviewed here for convenience.  Note that $\alpha$, $\beta$ and $\psi$
are defined in the hadronic rest frame, while $\theta$ is defined in
the $\tau$ rest frame.

%
\begin{figure}[t]
\begin{center}
\resizebox{4in}{!}{\includegraphics*{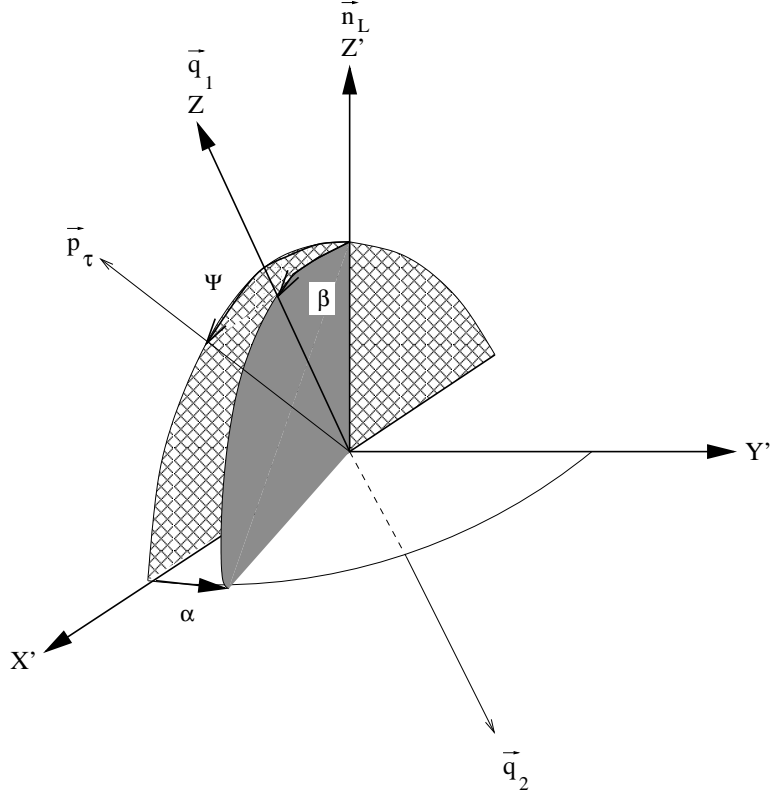}}
\caption{Coordinate system showing the angles for the decay $\tau\to
V\pi\nu$.  The momentum vectors shown are in the hadronic rest frame.
Note: this figure is very similar to one shown in
Ref.~\cite{Kpifinaltheory}.}
\label{fig:angles}
\end{center}
\end{figure}
%

Let the unit vector $\hat{n}_L$ denote the direction of the lab as
viewed in the hadronic rest frame and let $\hat{q}_1\equiv
\vec{q}_1/|\vec{q}_1|$, where $\vec{q}_1$ is the $V$'s momentum in the
hadronic rest frame.  Then we let $\beta$ denote the angle between
$\hat{n}_L$ and $\hat{q}_1$, as indicated in Fig.~\ref{fig:angles}.
The angles $\alpha$ and $\psi$ are also defined in
Fig.~\ref{fig:angles}.  As noted above, $\theta$ is defined in the
rest frame of the $\tau$.  In that frame, $\theta$ is the angle
between the direction of flight of the $\tau$ in the laboratory frame
and the direction of the hadrons.  Note that in some cases, even if
the $\tau$'s direction cannot be determined experimentally, the angles
$\theta$ and $\psi$ can both be measured since they are related in a
simple way to the energy $E_h$ of the hadronic system
\cite{deckermirkes}.

We assume here that the $\tau$'s have polarization $P$, aligned with
their direction of flight in the laboratory.  Then $s_\mu s^\mu=-P^2$.
A derivation of the $\tau$ spin four-vector $s^\mu$ in the hadronic
rest frame may be found in Appendix A of Ref.~\cite{KM1992}.  For
convenience, an expression is included in the Appendix of the present
work as well.

Since $\tau$ decays always involve a neutrino, it can be difficult in
practice to reconstruct the $\tau$'s momentum.  At this stage, we
assume that the $\tau$ direction is not known. Later, in our
discussion of triple-product CP asymmetries, we will assume that the
$\tau$ direction is in fact known. (Note that while the missing
neutrino certainly makes the determination of the $\tau$'s direction
difficult, there are experimental techniques that can be employed to
obtain the direction \cite{Kuhn:1993ra}.)

Employing the various definitions for the angles, and integrating over 
$\alpha$ (the unknown $\tau$ direction), we find~\cite{deckermirkes}
\bea
    d\Gamma = \frac{\left|\vec{q_1}\right|}{2m_\tau(4\pi)^3}
     \left(\frac{m_\tau^2-Q^2}{m_\tau^2}\right)
     \frac{dQ^2}{\sqrt{Q^2}}\frac{d\cos\theta}{2}\frac{d\cos\beta}{2}
     \left|{\cal A}\right|^2 \; .
     \label{eq:dGam}
\eea
The matrix element squared, $\left|{\cal A}\right|^2$, is proportional
to $L_{\mu\nu}\widetilde{H}^{\mu\nu}$ [see Eq.~(\ref{eq:Asq_2})].  We
write the physical polarization states of the $V$ in the hadronic rest
frame as
\bea
    \epsilon_1^\mu(\pm)=\frac{1}{\sqrt{2}}\left(0;\mp 1,i,0\right),~~~~
      \epsilon_1^\mu(0)=\left(-\frac{\left|q_1^z\right|}{m_V};
            0,0,-\frac{E_1}{m_V}\right) \; ,
    \label{eq:polnstates}
\eea
where the coordinate system is defined in Fig.~\ref{fig:angles}.  In
this basis, the hadronic tensor
$\widetilde{H}^{\mu\nu}=\widetilde{J}^\mu\left(\widetilde{J}^\nu\right)^\dagger$
is simplified considerably [see Eqs.~(\ref{eq:J0}) and (\ref{eq:Ji}),
and we sum over polarizations:
\bea
    \widetilde{H}^{\mu\nu} & \sim & \left(\begin{array}{cccc}
      \parallel & 0 & 0 & \parallel \\
      0 & \perp & \perp & 0 \\
      0 & \perp & \perp & 0 \\
      \parallel & 0 & 0 & \parallel \\
      \end{array}\right) \nn\\
    & \sim & \left(\begin{array}{cccc}
      |\widetilde{F}_4|^2 & 0 & 0 & \widetilde{F}_4 F_1^* , \widetilde{F}_4 F_2^*  \\
      0 & |F_1|^2, |F_3|^2 & {\rm Re}(F_1 F_3^*)  & 0 \\
      0 & {\rm Re}(F_1 F_3^*) & |F_1|^2, |F_3|^2 & 0 \\
      F_1\widetilde{F}_4^*, F_2\widetilde{F}_4^* & 0 & 0 & |F_1|^2, |F_2|^2, {\rm Re}( F_1 F_2^*)  \\
      \end{array}\right)  .
\eea
In the first matrix expression the non-zero elements are indicated
with the symbols ``$\parallel$'' and ``$\perp$''.  These denote the
longitudinal and transverse polarization states, respectively.  The
second matrix expression indicates the functional dependence on the
form factors $F_1(Q^2),\ldots,\widetilde{F}_4(Q^2)$.  Thus, for
example, $\widetilde{H}^{03}$ is only non-zero for
longitudinally-polarized $V$'s, and it depends on the
structure-function combinations $\widetilde{F}_4 F_1^*$ and
$\widetilde{F}_4 F_2^*$.

Recall that the NP contributions are contained in
$\widetilde{F}_4(Q^2)$, and that all CP-violating effects require the
interference of SM and NP amplitudes. We therefore see immediately
that any CP asymmetries will involve $\widetilde{H}^{00}$,
$\widetilde{H}^{03}$ or $\widetilde{H}^{30}$. This is consistent with
Eqs.~(\ref{eq:J0}) and (\ref{eq:Ji}), which note that only
$\widetilde{J}^0$ contains a NP contribution.

By defining symmetric and antisymmetric combinations of the various
elements of the tensors $L_{\mu\nu}$ and $\widetilde{H}^{\mu\nu}$, it
is possible to write the expression for
$L_{\mu\nu}\widetilde{H}^{\mu\nu}$ in a relatively simple manner.
Using the notation in Ref.~\cite{KM1992},\footnote{Note the
corrections to some of the relevant expressions in the Erratum to
Ref.~\cite{KM1992}.} we write
\bea
    L_{\mu\nu}\widetilde{H}^{\mu\nu} = \sum_X L_X W_X \; ,
   \label{eq:LXWX}
\eea
where $X=A,B,\ldots,I,SA,SB,\ldots,SG$ and where there is an implied
summation over polarization states.  Definitions and some expressions
for the relevant non-zero contributions are included in the Appendix.

We now examine three possible CP-violating signals: the rate
asymmetry, a polarization-dependent rate asymmetry (the polarization
is that of the $\tau$), and a triple-product asymmetry.

\subsection{Rate Asymmetry}

Integrating the differential width in Eq.~(\ref{eq:dGam}) over the
angular variables $\beta$ and $\theta$, and normalizing the result to
the width for $\tau\to e\nu\bar{\nu}$, we obtain
\bea
    \frac{d\Gamma}{\Gamma_e\,dQ^2}\simeq 
      \frac{\cos^2\theta_c}{2\left(Q^2\right)^{3/2}m_\tau^8}
        \left(m_\tau^2-Q^2\right)^2\left|q_1^z\right|
        \left[\left(2Q^2+m_\tau^2\right)\left(W_A+W_B\right)+3m_\tau^2 W_{SA}\right] \; ,
\eea
where $|q_1^z|$ is the magnitude of the momentum of the $V$ in the
hadronic rest frame (see the Appendix for an explicit expression) and
$\Gamma_e\simeq G_F^2m_\tau^5/(192\pi^3)$.  Note that $W_{SA}$
contains a Higgs-exchange contribution, but $W_A$ and $W_B$ do not
(see the Appendix). In particular,
\beq
W_{SA} = \widetilde{H}^{00} =
        \frac{Q^4\left(q_1^z\right)^2}{m_V^2}\left|\widetilde{F}_4\right|^2
        ~.  
\eeq

The rate asymmetry depends on the difference between the above
expression and the analogous one for the $\tau^+$ decay,
\bea
    A_{CP} = \frac{\Delta \Gamma}{\Gamma_{\textrm{\scriptsize{sum}}}} \; ,
     \label{eq:rateasym}
\eea
where $\Delta\Gamma$ is the difference of the widths for the process
and the anti-process (normalized to $\Gamma_e$),
\bea
    \Delta\Gamma &=& \int \left[\frac{d\Gamma}{\Gamma_e\,dQ^2}-
         \frac{d\overline{\Gamma}}{\Gamma_e\,dQ^2}\right] dQ^2 \nonumber \\
    & \simeq & \int 
        \frac{3\cos^2\theta_c}{2\left(Q^2\right)^{3/2}m_\tau^6}
        \left(m_\tau^2-Q^2\right)^2 \left|q_1^z\right|
        \left(W_{SA}-\overline{W}_{SA}\right)\, dQ^2 \; ,
   \label{eq:deltagam}
\eea
and $\Gamma_{\textrm{\scriptsize{sum}}}$ is the sum of the widths for
the process and the anti-process.

The expression for $\Delta\Gamma$ may be written explicitly in terms of strong
and weak phases.  Defining phases in the following manner,
\bea
    F_j(Q^2) & = & \left|F_j(Q^2)\right|\exp\left[i\delta_j(Q^2)\right] \; , ~~~~~~~(j=1,\ldots,4)\; ,
     \label{eq:F_j}\\
    f_H(Q^2) & = & \left|f_H(Q^2)\right|\exp\left[i\delta_H(Q^2)\right] \; , \\
    b & = & \left|b\right|\exp\left(i\phi_b\right) \; , \\
    a & = & \left|a\right|\exp\left(i\phi_a\right) \; ,
     \label{eq:b}
\eea
we have
\bea
    W_{SA}-\overline{W}_{SA} = 
       4 \frac{Q^4\left(q_1^z\right)^2}{m_V^2}\frac{\left|F_4f_Hb\right|}{m_\tau}
       \sin\left(\delta_4-\delta_H\right)\sin\phi_b 
\eea
for a vector meson (e.g.\ $V=\omega$). For an axial-vector meson
(e.g.\ $V=a_1$) the change $b\to a$ must be made. This expression may
be inserted in Eq.~(\ref{eq:deltagam}) to determine $\Delta\Gamma$.
The key point is that $\Delta\Gamma$ depends on the product of the NP
amplitude (proportional to $f_Hb$ or $f_Ha$) with the SM scalar
form factor $F_4$. But $F_4$ is expected to be very small,
regardless of whether $V$ is a vector or axial-vector meson. We
therefore conclude that the total rate asymmetries in $\tau$ decays
are small, even in the presence of NP.

\subsection{Polarization-dependent Rate Asymmetry}

The regular rate asymmetry defined in Eq.~(\ref{eq:rateasym}) depends
on $F_4$, and is likely to be very small. It is also possible to
define a rate asymmetry that depends on $f_HF_1^*$ and/or $f_HF_2^*$.
It turns out that to extract such cross-terms, one need only weight
the differential width in Eq.~(\ref{eq:dGam}) by $\cos\beta$ when
performing the integration over
$\cos\beta$.\footnote{Ref.~\cite{deckermirkes} shows some partial
results along these lines, but assumes that the scalar contributions
are zero.  Including the scalar contributions is straightforward.}
Performing the various integrals and forming a CP asymmetry, we
obtain
\bea
    A_{CP}^{\langle\cos\beta\rangle} = 
      \frac{\Delta \Gamma_{\langle\cos\beta\rangle}}{\Gamma_{\textrm{\scriptsize{sum}}}} \; ,
      \label{eq:acpbeta}
\eea
where $\Delta\Gamma_{\langle\cos\beta\rangle}$ is the difference of
the widths for the process and the anti-process:
\bea
    \Delta\Gamma_{\langle\cos\beta\rangle} &=& 
          \int \left[\frac{d\Gamma}{\Gamma_e\,dQ^2d\!\cos\beta}-
         \frac{d\overline{\Gamma}}{\Gamma_e\,dQ^2 d\!\cos\beta}\right] 
           dQ^2 \cos\beta \, d\!\cos\beta\nonumber \\
    & \simeq & -\int 
        \frac{\cos^2\theta_c}{2\left(Q^2\right)^{3/2}m_\tau^6}
        \left(m_\tau^2-Q^2\right)^2 \left|q_1^z\right|
        \left(W_{SF}-\overline{W}_{SF}\right)\, \rho(Q^2) dQ^2 \; ,
   \label{eq:deltagam_cosbeta}
\eea
with
\bea
    \rho(Q^2) \equiv \int \left(\cos\psi + P\cos\theta\cos\psi + 
       P\frac{\sqrt{Q^2}}{m_\tau}\sin\theta\sin\psi\right)\frac{d\cos\theta}{2} \; .
     \label{eq:rhoQ2}
\eea 
Note that $\psi$ is a function of $\theta$ (see the Appendix).

We have
\bea
W_{SF} & = & \widetilde{H}^{03}+\widetilde{H}^{30} \nn\\
& = & \frac{\left(Q^2\right)^{3/2}\left|q_1^z\right|}{m_V^2}
        \left[2\,\mbox{Re}\left(F_1\widetilde{F}_4^*\right)\left(\sqrt{Q^2}E_1\right)
          +4\,\mbox{Re}\left(F_2\widetilde{F}_4^*\right)\left|q_1^z\right|^2\right] ~,
\eea
where $E_1$ denotes the energy of the $V$ in the hadronic rest frame.
The form factors $F_1$ and $F_2$ are expected to be very small for a
vector meson (e.g.\ $V=\omega$) but are large for an axial-vector
meson (e.g.\ $V=a_1$). We therefore conclude that this asymmetry can
be sizeable for the decay $\tau \to a_1\pi \nu_\tau$.

\begin{figure}[t]
\begin{center}
\resizebox{4in}{!}{\includegraphics*{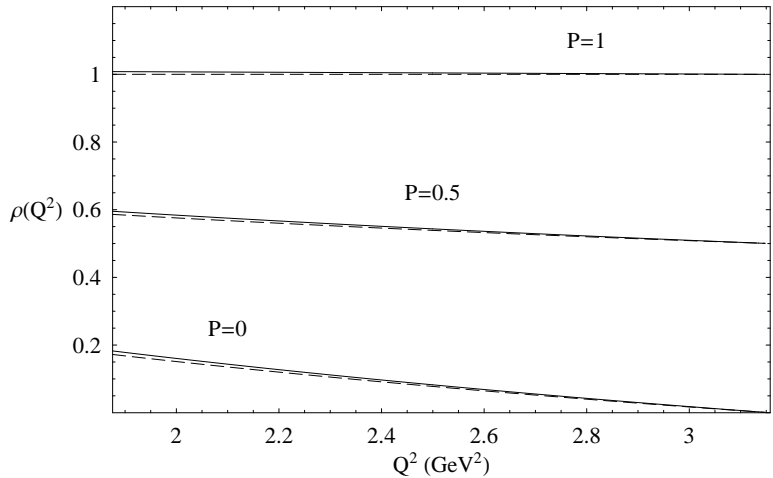}}
\caption{Plots of $\rho(Q^2)$ for various values of the $\tau$
polarization $P$.  The function $\rho(Q^2)$ is involved in the
polarization-dependent CP asymmetry and is defined in
Eq.~(\ref{eq:rhoQ2}).  The solid and dashed lines show the results for
$\sqrt{s}=m_{\Upsilon(4S)}$ and $\sqrt{s}=m_Z$, respectively.}
\label{fig:rhoQ2}
\end{center}
\end{figure}

Figure~\ref{fig:rhoQ2} shows plots of $\rho(Q^2)$ for a few parameter
choices, assuming that the $\tau$'s are produced in a symmetric
collider environment. We see that $\rho(Q^2)$ can be large, even if
the $\tau$'s are unpolarized ($P=0$).

%
%

Employing the definitions in Eqs.~(\ref{eq:F_j})-(\ref{eq:b}), and
assuming that the decay is $\tau \to a_1\pi \nu_\tau$, we find the
following expression for the difference $W_{SF}-\overline{W}_{SF}$,
\bea
    W_{SF}-\overline{W}_{SF} & = & 
       4 \frac{\left(Q^2\right)^{3/2}\left|q_1^z\right|}{m_{a_1}^2}
         \frac{\left|f_Ha\right|}{m_\tau} \nonumber \\
         & & \times \left[
           \left|F_1\right|\sin\left(\delta_1-\delta_H\right)\sqrt{Q^2}E_1
           + 2\left|F_2\right|\sin\left(\delta_2-\delta_H\right)\left(q_1^z\right)^2\right]
       \sin\phi_a ,~~~~~
       \label{eq:delwsf}
\eea
This expression may be inserted in Eq.~(\ref{eq:deltagam_cosbeta}) to
determine $\Delta\Gamma_{\langle\cos\beta\rangle}$. We stress that the
angle $\beta$ must be measured in this CP asymmetry. That is, in
contrast to the total rate asymmetry (which is very small), the
polarization-dependent rate asymmetry is in fact a {\it directional}
asymmetry.

As noted above, the four $\pi$'s in $\tau \to \pi \pi \pi \pi
\nu_\tau$ come primarily from the decay of $\omega \pi$ or $a_1 \pi$.
The polarization-dependent rate asymmetry of $\omega\pi$ is essentially
zero, so the (inclusive) measurement of this asymmetry in $\tau \to
\pi \pi \pi \pi \nu_\tau$ effectively probes the (exclusive)
contribution of $\tau \to a_1 \pi \nu_\tau$ and it appears that one
does not require that the $a_1$ be reconstructed. However, there is a
potential problem here. The angle $\beta$ is the angle between the
$a_1$ momentum in the hadronic rest frame and the direction of the lab
as viewed in this frame. In order to get $\beta$, one does in fact
have to reconstruct the $a_1$ to get its momentum. However, the width
of the $a_1$ is extremely large. As a consequence, it may not be
possible to identify unambiguously which three of the four pions came
from the $a_1$. If so, there will be an ambiguity in defining $\beta$.
There are two possible solutions to this problem. One is to exclude
events in which one cannot unambiguously reconstruct the $a_1$. This
may severely reduce the number of events. The second is to include all
possible reconstructions of the $a_1$ \cite{4pi}. This may severely
reduce the signal. We assume that the experimentalists will find the
optimal solution.

We now present a brief numerical analysis of the
polarization-dependent asymmetry for $\tau \to a_1 \pi \nu_\tau$.  The
polarization asymmetry depends on cross terms involving the NP form
factor $f_H$ with the SM form factors $F_1$ and $F_2$ [see
Eq.~(\ref{eq:delwsf})].  The form factors $F_1$ and $F_2$ are expected
to dominate in the calculation of the width for $\tau \to a_1 \pi
\nu_\tau$, but at this point the relative sizes of the two are not
known.  To perform a few estimates of the polarization asymmetry, we
make the simplifying assumption that either $F_1=0$ or
$F_2=0$.\footnote{Although the real picture is expected to correspond
to some intermediate situation, this seems to be a reasonable way to
proceed in order to make model-independent predictions.}  In addition,
since the $Q^2$ dependence of the form factors is theoretically
unknown, we have repeated the calculations for two different sets of
assumptions in each case.

In the first scenario, we assume that $F_1$ (or $F_2$) receives
contributions only from the $\rho(770)$; in the second, we assume that
both the $\rho(770)$ and $\rho(1450)$ resonances contribute.  The form
factors thus have the form shown for
$F_3^{\textrm{\scriptsize{pole}}}$ in Eq.~(\ref{omega_res}), but with
$g_{\rho\omega\pi}/\gamma_\rho$ replaced by a constant that is
determined by the experimental rate for ${\cal B}(\tau \to a_1 \pi
\nu_\tau)$.\footnote{In the first case we set $A_1=0$ (i.e., $\rho$
resonance only).  In the second case we somewhat arbitrarily take
$A_1=-0.24$, which is the value we also use in Sec.~\ref{sec:3.4} when
estimating the triple product asymmetry in $\tau \to \omega \pi
\nu_\tau$.  In that case the choice is well-motivated~\cite{4pi}.}
The CLEO collaboration has performed several fits to determine the
branching ratio for $\tau \to a_1 \pi \nu_\tau$.  In one of the better
fits, the collaboration obtains $R_{a_1 \pi} = {\cal B}(\tau^- \to
{(a_1 \pi)}^- \nu_\tau) / {\cal B}(\tau^- \to 2 \pi^- \pi^+ \pi^0
\nu_\tau)=0.49\pm0.02\pm0.02$ (Model 3 in Table IV of
Ref.~\cite{4pi}).  There is a slight subtlety in that this ratio
includes contributions from both $\tau^- \to
a_1^0 \pi^- \nu_\tau$ and $\tau^- \to
a_1^- \pi^0 \nu_\tau$.  Furthermore, the $a_1^0$ can be formed
in two different ways, while the $a_1^-$ can only be formed in one way.
The experimental analysis includes all three possibilities and even
allows them, in principle, to interfere.  For our analysis we consider the
$a_1^0$ case and, as an approximation, use the central value 
$R_{a_1^0 \pi^-}\simeq (2/3)\times R_{a_1 \pi}\simeq 0.327$
to determine $F_1$ (or $F_2$).  We then assume that a NP
contribution could reasonably be assumed to be ``hidden'' in the
remaining uncertainty.  We thus tune the NP combination $\left|a
f_H\right|$ such that the NP contribution to $R_{a_1^0 \pi^-}$ is equal to
approximately $0.019$ (i.e., we multiply the uncertainty by $2/3$ as well).  
This procedure yields the value
$\left|af_H\right|\simeq 6.8$.  (Note that $f_H$ is assumed to be a
constant with zero strong phase.)  Having fixed $F_1$ (or $F_2$) in
this manner, and setting $P$=1 in Eq.~(\ref{eq:rhoQ2}), we proceed to
compute the polarization asymmetry defined in Eq.~(\ref{eq:acpbeta}).

If $F_{1,2}$ are assumed to be dominated by the $\rho(770)$ resonance
we obtain quite small asymmetries -- of order $\sim 1\%$ for
$\left|\sin\phi_a\right|=1$ -- regardless of which form factor has
been set to zero.  This fact is not surprising in this case: since the
$\rho(770)$ cannot go on shell in $\tau \to a_1 \pi \nu_\tau$, the
factors $\sin\left(\delta_{1,2}-\delta_H\right)$ in
Eq.~(\ref{eq:delwsf}) are small (recall that we have assumed
$\delta_H=0$).  Larger asymmetries are obtained by assuming that both
the $\rho$ and $\rho^\prime$ contribute to $F_{1,2}$, since the
$\rho^\prime$ can go on-shell, leading to appreciable strong phase
effects.  In this case asymmetries of order $15 \%$ ($7.5 \%$) can be
obtained when $F_1$ ($F_2$) is set to zero (again assuming that the
strong phase associated with the NP contribution, $\delta_H$, is set
to zero).

\subsection{Triple-product Asymmetry}
\label{sec:3.4}

If the momentum of the $\tau$ can be determined experimentally, it is
possible to construct a triple-product CP asymmetry.  In this
subsection we will assume that the $\tau$ is unpolarized ($P=0$), but
that the polarization of the $V$ is measured.  Furthermore, we will
assume that the polarization of the $V$ is neither purely transverse,
nor purely longitudinal.  As we shall see, this assumption will allow
the NP contribution to interfere with the SM form factor $F_3$,
allowing for possibly observable effects for $V=\omega$ and $V=\rho$.

There are in principle two possible sources of triple products in
hadronic $\tau$ decays, assuming that the $\tau$'s are unpolarized.
The first possible source arises from the leptonic tensor
$L_{\mu\nu}$, which contains a term proportional to
$i\varepsilon_{\alpha\mu\beta\nu}l^{\prime\alpha}l^\beta$.  This term
is contracted with
$\widetilde{H}^{\mu\nu}=\widetilde{J}^\mu\left(\widetilde{J}^{\nu}\right)^\dagger$,
which only contains NP contributions if $\mu$ and/or $\nu$ is equal to
zero [see Eqs.~(\ref{eq:J0}) and (\ref{eq:Ji})].  It is then evident
that the $i\varepsilon_{\alpha\mu\beta\nu}l^{\prime\alpha}l^\beta$
term in $L_{\mu\nu}$ cannot lead to a CP-odd TP asymmetry.  (Note that
$i\varepsilon_{i0jk}l^{\prime i}l^j=0$ because $\vec{l}^\prime =
\vec{l}$ in the hadronic rest frame.)

The only other source of a CP-odd TP asymmetry is found in the
hadronic tensor; the triple product arises from a Levi-Civita tensor
term in $\widetilde{J}^\mu$ involving the $F_3$ term. This $F_3$ can
be large if $V$ is a vector meson: $V=\omega$, $\rho$. The
CP-violating TP asymmetry can be sizeable for the decays
$\tau\to\omega\pi\nu_\tau$ and $\tau\to\rho\pi\nu_\tau$.

The TP involves the $V$ momentum, so that it is necessary to
reconstruct the $V$. This makes the decay $\tau\to\rho\pi\nu_\tau$
problematic. The width of the $\rho$ is very large and one runs into a
similar problem as that described in Sec.~3.3. In the decay
$\tau\to\rho\pi\nu_\tau \to 3\pi \nu_\tau$, it is not possible to
identify unambiguously which two of the three pions came from the
$\rho$. It is therefore very difficult to reconstruct the $\rho$ and,
as a consequence, difficult to measure the TP in these decays.

The decay $\tau\to\omega\pi\nu_\tau$ does not suffer the same
problem. The $\omega$ is quite narrow, so it can be fairly easily
reconstructed. For this reason, in the measurement of the TP
asymmetry, we exclude $\tau\to\rho\pi\nu_\tau$, concentrating only on
$\tau\to\omega\pi\nu_\tau$. (Note that this implies that there are no
clean CP-violating signals in $\tau\to 3\pi \nu_\tau$.)

Keeping only the terms that can lead to a CP-odd TP asymmetry, and
averaging over the spin states of the initial $\tau$, we have
\bea
    \left. L_{\mu\nu}\widetilde{H}^{\mu\nu} \right|_{\textrm{\scriptsize{TP}}}
    & = & \left[-L^{0i}\widetilde{J}^0\left(J^i\right)^\dagger
              -L^{i0}J^i\left(\widetilde{J}^0\right)^\dagger
                  \right]_{\textrm{\scriptsize{TP}}} \\
    & = & \frac{8m_\tau Q^2}{E_1} \mbox{Im}\left(bf_HF_3^*\right) 
           \left(\vec{\epsilon}_1\cdot\vec{q}_1\right)
          \vec{\epsilon}_1\cdot \left(\vec{l}\times\vec{q}_1\right) \; .
      \label{eq:LH_TP}
\eea
In the second line we have adopted a real basis for the $\omega$
polarization tensor and have used the relation $\epsilon_1\cdot
q_2=\sqrt{Q^2}\,\vec{\epsilon}_1\cdot\vec{q}_1/E_1$, which is valid in
the hadronic rest frame.

Since we now assume that the $\tau$ direction is known, we can
simplify our coordinate system somewhat compared to that shown in
Fig.~\ref{fig:angles}.  We define $\zeta$ to be the angle between
$\vec{l}$ and $\vec{q}_1$ ($\vec{l}$ is the momentum of the $\tau$;
$\vec{q}_1$ is the momentum of the $V$); that is,
\bea
    \vec{l}\cdot\vec{q}_1 = \left|\vec{l}\,\right|\left|\vec{q}_1\right| \cos\zeta \; .
\eea
It is then convenient to define three direction vectors as follows,
\begin{eqnarray}
  \vec{n}_1 = \frac{\vec{q}_1}{\left|\vec{q}_1\, \right|}\frac{m_\omega}{E_1} \; , ~~~~
  \vec{n}_2 = \frac{\vec{l}\times \vec{q}_1}
     {\left|\vec{l}\times \vec{q}_1\right|} = \frac{\vec{l}\times \vec{q}_1}
             {\left|\vec{l}\,\right|\left|\vec{q}_1\right| \sin\zeta}\; , ~~~~
  \vec{n}_3 = \frac{\left(\vec{l}\times \vec{q}_1\right)\times\vec{q}_1}
     {\left|\left(\vec{l}\times \vec{q}_1\right)\times\vec{q}_1\, \right|} \; , 
\end{eqnarray}
where all quantities are expressed in the hadronic rest frame.  The
peculiar normalization of $\vec{n}_1$ is associated with the boost
from the $\omega$'s rest frame to the hadronic rest frame.  With the
above definitions, the condition $\epsilon_1^2=-1$ becomes 
\begin{eqnarray}
  \left(\vec{\epsilon}_1\cdot\vec{n}_1\right)^2 + 
    \left(\vec{\epsilon}_1\cdot\vec{n}_2\right)^2 + 
    \left(\vec{\epsilon}_1\cdot\vec{n}_3\right)^2 =1 \; .
\end{eqnarray}

Rewriting Eq.~(\ref{eq:LH_TP}) in terms of the various direction vectors, we have
\bea
    \left. L_{\mu\nu}\widetilde{H}^{\mu\nu} \right|_{\textrm{\scriptsize{TP}}}
    & = & \frac{8m_\tau \left|q_1^z\right|Q^2}{m_\omega} \sin\zeta \, \mbox{Im}\left(bf_HF_3^*\right) 
           \left(\vec{\epsilon}_1\cdot\vec{n}_1\right)
           \left(\vec{\epsilon}_1\cdot\vec{n}_2\right)\; .
\eea
Substituting this expression into Eq.~(\ref{eq:Asq_2}) and calculating
the differential width, we find,
\bea
    \left.\frac{d\Gamma}{\Gamma_e\,dQ^2}\right|_{\textrm{\scriptsize{TP}}}= 
      \frac{3\pi\cos^2\theta_c}{4\,m_\omega m_\tau^6\sqrt{Q^2}}
        \left(m_\tau^2-Q^2\right)^2\left|q_1^z\right|^3
        \, \mbox{Im}\left(bf_HF_3^*\right) 
           \left(\vec{\epsilon}_1\cdot\vec{n}_1\right)
           \left(\vec{\epsilon}_1\cdot\vec{n}_2\right)\; ,
         \label{eq:dgam_TP}
\eea
in which we have integrated over the angle $\zeta$.

It is now possible to define a CP asymmetry based on Eq.~(\ref{eq:dgam_TP}),
\bea
    A_{CP}^{\textrm{\scriptsize{TP}}} = 
      \frac{\Delta \Gamma_{\textrm{\scriptsize{TP}}}}{\Gamma_{\textrm{\scriptsize{sum}}}} \; ,
\eea
where $\Delta\Gamma_{\textrm{\scriptsize{TP}}}$ is the difference of
the widths for the process and the anti-process,
\bea
    \Delta\Gamma_{\textrm{\scriptsize{TP}}} &=& 
          \int \left[\frac{d\Gamma}{\Gamma_e\,dQ^2}-
         \frac{d\overline{\Gamma}}{\Gamma_e\,dQ^2}\right]_{\textrm{\scriptsize{TP}}}
           dQ^2 \nonumber \\
           & & \nonumber \\
    & \simeq & \left(\vec{\epsilon}_1\cdot\vec{n}_1\right)
           \left(\vec{\epsilon}_1\cdot\vec{n}_2\right) 
           \int 
           \frac{3\pi\cos^2\theta_c}{2\,m_\omega m_\tau^6\sqrt{Q^2}}
        \left(m_\tau^2-Q^2\right)^2\left|q_1^z\right|^3 \nonumber \\
        & & ~~~~~~~~~~~~~~~~~~~~\times \left|bf_HF_3\right| 
            \cos\left(\delta_3-\delta_H\right)\sin\phi_b \, dQ^2 .~~~
   \label{eq:deltagam_TP}
\eea
This CP-odd TP asymmetry depends on the form-factor combination
$f_HF_3$, which is the most favourable combination possible, since
$F_3$ is known to dominate over the other form factors in
Eq.~(\ref{Wamp}) (for the decay $\tau\to\omega\pi\nu_\tau$).
Furthermore, as noted above, the asymmetry vanishes if the $\omega$'s
polarization is either purely longitudinal
($\vec{\epsilon}_1\cdot\vec{n}_2=0$) or purely transverse
($\vec{\epsilon}_1\cdot\vec{n}_1=0$).  Finally, note that the TP
asymmetry depends on the cosine of the strong phase difference, so it
can be non-zero even if there are no strong phases.

The TP asymmetry defined above is written explicitly in terms of
projections of the polarization vector onto the vectors $\vec{n}_1$
and $\vec{n}_2$.  A similar approach was followed in
Refs.~\cite{wkn1,wkn2}.  In that case the authors constructed T-odd
observables in $B\to D^* \ell \nu$ using the polarization of the
$D^*$.  In Appendix B of Ref.~\cite{wkn2}, the authors showed that
there was a direct connection between the asymmetries written in terms
of polarization projections and more ``physical'' asymmetries
constructed using the momenta of the $D^*$ decay products.

We conclude this section by presenting a numerical investigation of
the size of the TP asymmetry in the decay
$\tau\to\omega\pi\nu_\tau$. This asymmetry is easier to study than the
one discussed in the previous section, since the SM decay amplitude is
dominated by a single SM form factor, $F_3$.  For our numerical
estimate of the TP asymmetry, we use the expression for $F_3$ given in
Eq.~(\ref{omega_res}), taking $A_1=-0.24$,
$g_{\rho\omega\pi}=16.1$~GeV$^{-1}$ and $\gamma_\rho\simeq 4.95$
(Model 2 in Table I of Ref.~\cite{4pi}).  This yields $R_{\omega \pi}
= {\cal B}(\tau^- \to {(\omega \pi)}^- \nu_\tau) / {\cal B}(\tau^- \to
2 \pi^- \pi^+ \pi^0 \nu_\tau)\simeq 0.38$, in agreement with Model 3
of Table IV of Ref.~\cite{4pi}, which had obtained the value
$R_{\omega \pi}=0.38\pm 0.02\pm 0.01$.  Assuming that the NP
contribution can at most be responsible for the remaining
uncertainty in $R_{\omega \pi}$, we tune $\left|bf_H\right|$
such that its contribution to $R_{\omega \pi}$ is approximately
$0.022$.  This procedure yields $\left|bf_H\right|\simeq 0.98$.
Finally, we have computed the asymmetry using this value and assuming
that the strong phase $\delta_H=0$ and that
$\left|\sin\phi_b\right|=1$.  The resulting TP asymmetry has a magnitude
of order $30\%$ multiplied by
$\left(\vec{\epsilon}_1\cdot\vec{n}_1\right)
\left(\vec{\epsilon}_1\cdot\vec{n}_2\right)$

\section{Conclusions}

In this paper we have re-examined CP violation in $\Delta S = 0$
hadronic $\tau$ decays. Any CP-violating effect requires the
interference of (at least) two amplitudes with a relative weak phase
(and often a relative strong phase). The standard-model contribution
involves (essentially) only a virtual $W$ and constitutes one
amplitude. The second amplitude must come from new physics (NP).  We
assume that the NP consists of a charged Higgs boson. This is a fairly
general assumption.

However, in many models, the coupling of the charged Higgs boson to
first-generation final states is proportional to first-generation
masses, and is negligible. In order to obtain a measurable
CP-violating effect in $\tau$ decays, we must examine non-``standard''
NP, and consider only the case in which the charged-Higgs coupling to
first-generation particles is large. This is not usual, but if CP
violation is seen in $\tau$ decays, it points clearly to a different
type of NP.

The decays examined are essentially $\tau \to N\pi \nu_\tau$, with
$N=2,3,4$. We have shown that the charged Higgs $H$ does not couple to
two pions, and so the decay $\tau \to \pi \pi \nu_\tau$ will not
exhibit CP violation. However, $H$ does couple to $3\pi$ and $4\pi$,
so that CP violation could be seen in $\tau \to N\pi \nu_\tau$
($N=3,4$). In addition to non-resonant contributions, these decays
receive resonant contributions from intermediate particles such as
$\omega$, $\rho$ and $a_1$. These resonant contributions involve a
smaller number of final particles, and are dominant. We therefore
consider only the resonant contributions to $\Delta S = 0$ hadronic
$\tau$ decays.

For the decay $\tau \to V\pi \nu_\tau$ ($V$ is a vector or
axial-vector meson), the SM hadronic current involves four form
factors $F_{1{\hbox{-}}4}(Q^2)$. If $V$ is a vector meson ($\omega$,
$\rho$), the contribution proportional to $F_3$ dominates. The $F_1$
and $F_2$ terms violate $G$-parity, and there is no evidence for such
second-class currents. Equally, there is no evidence for a non-zero
$F_4$ term, which is scalar. If $V$ is an axial-vector meson ($a_1$),
$F_1$/$F_2$ and $F_3$ change roles. Thus, in the decay $\tau \to
a_1\pi \nu_\tau$, the contributions proportional to $F_1$ and $F_2$
dominate, while the $F_3$ term violates $G$-parity. For vector mesons,
the NP hadronic current is proportional to $b f_{\sss H}$, where
$f_{\sss H}$ is a form factor and $b$ is assumed to have a non-zero
weak phase ($b$ is the pseudoscalar coupling of the charged Higgs).
For axial-vector mesons, one makes the substitution $b\to a$, where
$a$ is the scalar coupling of the charged Higgs.

We have examined three potential signals of CP violation: an overall
rate asymmetry, a polarization-dependent rate asymmetry (the
polarization is that of the $\tau$), and a triple-product
asymmetry. It is found that the overall rate asymmetry is proportional
to $|f_{\sss H} F_4|$ and is tiny for all hadronic $\tau$ decays.

The polarization-dependent rate asymmetry is proportional to $|f_{\sss
H} F_1|$ and $|f_{\sss H} F_2|$. It is therefore sizeable only for
$\tau \to a_1\pi \nu_\tau$. However, the width of the $a_1$ is
extremely large, which can lead to ambiguities in reconstructing the
direction of the $a_1$ momentum. If this problem can be addressed, the
polarization-dependent rate asymmetry in $\tau \to a_1\pi \nu_\tau \to
4\pi \nu_\tau$ is a good signal of CP violation, with values
of order $15\%$ currently allowed by experimental data.

Finally, the triple-product asymmetry is proportional to $|f_{\sss H}
F_3|$, and can be sizeable for $\tau \to \omega\pi \nu_\tau$ and $\tau
\to \rho\pi \nu_\tau$. However, the $\rho$ width is very large, which
can lead to difficulties in reconstructing the $\rho$ from $\tau \to
3\pi \nu_\tau$. The $\omega$ is quite narrow and does not suffer
similar problems. For this reason we consider only the decay $\tau \to
\omega\pi \nu_\tau$ in the calculation of the TP asymmetry.  Note that
this means that there are no clean signals of CP violation in $\tau
\to 3\pi \nu_\tau$.  Our analysis indicates that the current
experimental data allow a TP asymmetry for $\tau \to \omega\pi
\nu_\tau$ as large as $30\%$ multiplied by
$\left(\vec{\epsilon}_1\cdot\vec{n}_1\right)
\left(\vec{\epsilon}_1\cdot\vec{n}_2\right)$.

Note also that this triple product is $\epsilon_\omega \cdot (p_\omega
\times p_\tau)$. In order to measure this TP, one needs the spin of
the $\omega$ ($\epsilon_\omega$), which in turn means that the
direction of the $\tau$ must be known.

\bigskip
\noindent
{\bf Acknowledgments}:
We thank the following people for helpful communications: S. Banerjee,
T. Browder, H. Hayashii, B. Kowalewski, M. Roney and B. Viaud. A.D.
and K.K. thank the Physics Departments of the University of Toronto
and the Universit\'e de Montr\'{e}al, respectively, for their
hospitality during the initial stages of this work. This work was
financially supported by NSERC of Canada. The work of K.K. was
supported in part by the U.S.\ National Science Foundation under
Grants PHY--0301964 and PHY--0601103.


\begin{appendix}
\section{Appendix}

This section contains the definitions for the relevant $L_X$ and $W_X$
in Eq.~(\ref{eq:LXWX}) as well as explicit expressions for several
other quantities, such as $\left|q_1^z\right|$, $\cos\theta$ and
$\cos\psi$.

\subsection{\boldmath Expressions for $L_X$ and $W_X$.}

In the polarization basis indicated in Eqs.~(\ref{eq:polnstates}) many
of the elements of the hadronic tensor $\widetilde{H}^{\mu\nu}$ are
zero.  Summing over the polarization states, we find
\bea
    L_{\mu\nu}\widetilde{H}^{\mu\nu} = L_AW_A +L_BW_B +L_EW_E +
      L_{SA}W_{SA}+L_{SF}W_{SF} +L_{SG}W_{SG}  \; , 
\eea
where~\cite{KM1992},
\bea
    \begin{array}{rclrcl}
    L_A & = & \frac{1}{2}\left(L^{11}+L^{22}\right) \; , 
       &~~~~~~~~W_A & = & \widetilde{H}^{11}+\widetilde{H}^{22} \; ,\\
       &&&& \nonumber\\
    L_B & = & L^{33} \; , &  W_B & = & \widetilde{H}^{33}  \; ,\\
       &&&& \nonumber\\
    L_E & = & \frac{i}{2}\left(L^{12}-L^{21}\right) \; , 
       & W_E & = &-i\left(\widetilde{H}^{12}-\widetilde{H}^{21}\right)  \; , \\
       &&&& \nonumber\\
    L_{SA} & = & L^{00} \; , & W_{SA} & = & \widetilde{H}^{00}  \; ,\\
       &&&& \nonumber\\
    L_{SF} & = & -\frac{1}{2}\left(L^{03}+L^{30}\right)\; , 
       & W_{SF} & = & \widetilde{H}^{03}+\widetilde{H}^{30} \; ,\\
       &&&& \nonumber\\
    L_{SG} & = & -\frac{i}{2}\left(L^{03}-L^{30}\right)  \; , 
       & W_{SG} & = & -i\left(\widetilde{H}^{03}-\widetilde{H}^{30}\right) \; .\\
    \end{array}
\eea
Explicit expressions for these quantities may be found in
Ref.~\cite{deckermirkes} (recalling that one needs to make the
replacement $F_4\to\widetilde{F}_4$).  A few expressions that are
useful for our purposes are the following, 
\bea
    W_A & = &
       2Q^2\left[\left(q_1^z\right)^2\left|F_3\right|^2+Q^2\left|F_1\right|^2\right] \; ,\\
    W_B & = & 
       \frac{Q^2}{m_V^2}\left[Q^2E_1^2\left|F_1\right|^2+4\sqrt{Q^2}E_1\left(q_1^z\right)^2\mbox{Re}\left(F_1F_2^*\right)
         + 4\left(q_1^z\right)^4\left|F_2\right|^2\right] \; ,\\
    W_{SA} & = & \frac{Q^4\left(q_1^z\right)^2}{m_V^2}\left|\widetilde{F}_4\right|^2
        \; ,\\
    W_{SF} & = & \frac{\left(Q^2\right)^{3/2}\left|q_1^z\right|}{m_V^2}
        \left[2\,\mbox{Re}\left(F_1\widetilde{F}_4^*\right)\left(\sqrt{Q^2}E_1\right)
          +4\,\mbox{Re}\left(F_2\widetilde{F}_4^*\right)\left|q_1^z\right|^2\right]
        \; ,\\
    W_{SG} & = & \frac{\left(Q^2\right)^{3/2}\left|q_1^z\right|}{m_V^2}
        \left[2\,\mbox{Im}\left(F_1\widetilde{F}_4^*\right)\left(\sqrt{Q^2}E_1\right)
          +4\,\mbox{Im}\left(F_2\widetilde{F}_4^*\right)\left|q_1^z\right|^2\right]
        \; .
\eea
Note that only $W_A$ is non-zero in the limit that the $F_3$ structure
function dominates the current $\widetilde{J}^\mu$, as is the case for
$\tau \to \omega \pi \nu_\tau$.  Furthermore, as noted in the text,
the form factor $\widetilde{F}_4$, which contains the Higgs
contribution, only appears in combination with $F_1$, $F_2$ and
itself, and not with the dominant $F_3$ term.  The combination
$F_3\widetilde{F}_4^*$ can arise if the $V$'s polarization is taken to
have non-zero projections in both the transverse and longitudinal
directions.  Such is the case for the triple-product CP asymmetry
constructed in the text.

\subsection{\boldmath Expressions for some kinematical quantities.}

In the hadronic rest frame the magnitude of the momentum of the $V$ is
given by
\bea
    \left|q_1^z\right| = 
      \frac{1}{2\sqrt{Q^2}}\, \lambda^{1/2}\!\left(Q^2,m_V^2,m_\pi^2\right) \; ,
\eea
where $\lambda(x,y,z)\equiv x^2+y^2+z^2-2(xy + xz + yz)$.  Then the
energy of the $V$ is simply
$E_1=\sqrt{\left(q_1^z\right)^2+m_V^2}$.

It is also possible to derive a relation between the angles $\theta$
and $\psi$.  Assuming that the $\tau$'s are pair-produced at a
symmetric collider, these angles may be expressed as follows~\cite{deckermirkes},
\bea
    \cos\theta & = & \frac{2xm_\tau^2-m_\tau^2-Q^2}
               {\left(m_\tau^2-Q^2\right)\left(1-4m_\tau^2/s\right)^{1/2}} \; ,
           \label{eq:costheta} \\
    \cos\psi & = & \frac{x\left(m_\tau^2+Q^2\right)-2Q^2}
               {\left(m_\tau^2-Q^2\right)\left(x^2-4Q^2/s\right)^{1/2}} \; ,   
           \label{eq:cospsi}   
\eea
where $s=4 E_{\textrm{\scriptsize{beam}}}^2=4E_\tau^2$ and $x=2
E_h/\sqrt{s}$, with $E_h$ being the hadron energy in the lab frame.
To determine the relation between $\psi$ and $\theta$ explicitly, one
can solve Eq.~(\ref{eq:costheta}) for $x$ in terms of $\cos\theta$ and
then substitute the result into Eq.~(\ref{eq:cospsi}).  This allows
one, for example, to perform the integration in Eq.~(\ref{eq:rhoQ2}).
Similar expressions could be derived in the case of an asymmetric collider,
assuming one knew the correct boost factor.

In determining the polarization-dependent CP asymmetry, we have
assumed the incident $\tau$'s to have polarization $P$ as viewed from
the lab.  A derivation of the $\tau$ spin four-vector in the hadronic
rest frame may be found in Appendix A of Ref.~\cite{KM1992}.  For
convenience, we list the result here as well.  In the $S^\prime$ frame
(see Fig.~\ref{fig:angles}),
\bea
    \left(s^\mu\right)^{S^\prime} & = &
          P\left(-\frac{\left|\vec{l}\right|}{m_\tau}\cos\theta,
          -\frac{l_0}{m_\tau}\cos\theta\sin\psi+\sin\theta\cos\psi,0, \right. \nonumber \\
          & & \left.
          ~~~~~~~~~~~~~~~~~~~~~
          -\frac{l_0}{m_\tau}\cos\theta\cos\psi-\sin\theta\sin\psi\right) \; ,
\eea
where $l_0$ and $\left|\vec{l}\,\right|$ refer to the energy and
momentum of the $\tau$ in the hadronic rest frame.  To obtain the
expression in the $S$ frame, one needs to perform an orthogonal
rotation on the spatial components of $\left(s^\mu\right)^{S^\prime}$
using the angles $\alpha$ and $\beta$ indicated in
Fig.~\ref{fig:angles}.

\end{appendix}



\begin{thebibliography}{99}

\bibitem{pdg} S.~Eidelman {\it et al.}  [Particle Data Group], Phys.\
Lett.\ B {\bf 592}, 1 (2004).

\bibitem{CLS} S.~Y.~Choi, J.~Lee and J.~Song,
Phys.\ Lett.\ B {\bf 437}, 191 (1998);

\bibitem{Kpifinaltheory} J.~H.~Kuhn and E.~Mirkes,
Phys.\ Lett.\ B {\bf 398}, 407 (1997).

\bibitem{3pifirst} S.~Y.~Choi, K.~Hagiwara and M.~Tanabashi,
Phys.\ Rev.\ D {\bf 52}, 1614 (1995).

\bibitem{3pifinal} Y.~S.~Tsai,
arXiv:hep-ph/9801274.

\bibitem{decker87}
R.~Decker,
Z.\ Phys.\ C {\bf 36}, 487 (1987).

\bibitem{deckermirkes}
R.~Decker and E.~Mirkes,
Z.\ Phys.\ C {\bf 57}, 495 (1993).

\bibitem{wise}
H.~Davoudiasl and M.~B.~Wise,
Phys.\ Rev.\ D {\bf 53}, 2523 (1996).

\bibitem{KM1992} J.~H.~Kuhn and E.~Mirkes, Z.\ Phys.\ C {\bf 56}, 661
  (1992) [Erratum-ibid.\ C {\bf 67}, 364 (1995)].

\bibitem{KM1997} J.~H.~Kuhn and E.~Mirkes, Phys.\ Lett.\ B {\bf 398},
407 (1997).

\bibitem{Kuhn:1993ra} J.~H.~Kuhn, Phys.\ Lett.\ B {\bf 313}, 458
(1993).

\bibitem{CPVpipiexpt} P.~Avery {\it et al.}  [CLEO Collaboration],
Phys.\ Rev.\ D {\bf 64}, 092005 (2001).

\bibitem{BS} I.~I.~Bigi and A.~I.~Sanda, Phys.\ Lett.\ B {\bf 625}, 47
(2005).

\bibitem{CPVKpiexpt} G.~Bonvicini {\it et al.}  [CLEO
Collaboration],
Phys.\ Rev.\ Lett.\ {\bf 88}, 111803 (2002);
S.~Anderson {\it et al.}  [CLEO Collaboration],
Phys.\ Rev.\ Lett.\  {\bf 81}, 3823 (1998).

\bibitem{4pi} K.~W.~Edwards {\it et al.}  [CLEO Collaboration],
Phys.\ Rev.\ D {\bf 61}, 072003 (2000).

\bibitem{aleph}
P.~Bourdon,
Nucl.\ Phys.\ Proc.\ Suppl.\  {\bf 40}, 203 (1995).

\bibitem{aleph1997}
D.~Buskulic {\it et al.}  [ALEPH Collaboration],
Z.\ Phys.\ C {\bf 74}, 263 (1997).

\bibitem{wkn1} G.~H.~Wu, K.~Kiers and J.~N.~Ng, Phys.\ Lett.\ B {\bf
  402}, 159 (1997).

\bibitem{wkn2} G.~H.~Wu, K.~Kiers and J.~N.~Ng, Phys.\ Rev.\ D {\bf
  56}, 5413 (1997).  

\end{thebibliography}
\end{document}